\documentclass[a4paper,12pt]{article}
\usepackage{amssymb,amsmath,amsthm,mathrsfs,mathtools}
\usepackage{caption}
\usepackage{cite}
\usepackage{hyperref}
\usepackage[affil-it]{authblk}
\usepackage{tikz,pgfplots,tkz-euclide}
\usetikzlibrary{arrows,positioning,calc,fadings,decorations.pathreplacing,decorations}
\def\R{\mathbf{R}}

\def\K{\mathcal{K}}
\newcommand\G[1]{\Gamma(#1)}
\newcommand\ads{\ensuremath{\mathcal{M}_{\text{AdS}}}}
\newcommand\ds{\ensuremath{\mathcal{M}_{\text{dS}}}}
\newcommand\eq[1]{(\ref{#1})}
\newcommand\C{\mathscr{C}}

\def\pa{\partial}

\newcommand\rt{\longrightarrow}

\title{Holography in de Sitter and anti-de Sitter Spaces and Gel'fand Graev
Radon transform}
\author[1]{Samrat Bhowmick\thanks{email: bhowmicksamrat@gmail.com}~} 
\author[1]{Koushik Ray\thanks{email: koushik@iacs.res.in}}
\author[1,2]{Siddhartha Sen\thanks{email: sen1941@gmail.com}}
\affil[1]{Indian Association for the Cultivation of Science,\authorcr
Calcutta 700 032. India.}
\affil[2]{CRANN, Trinity College Dublin, Dublin -- 2, Ireland}
\date{}
\begin{document}
\maketitle
\begin{abstract}
\noindent
Bulk reconstruction formulas similar to HKLL are obtained for de Sitter and
anti-de Sitter spaces as the inverse Gel'fand Graev Radon transform. 
While these generalize our previous result on the Euclidean anti-de Sitter
space, their validity in here is restricted  only to odd dimensions in both
instances. The exact Wightman function for the de Sitter space is then
derived.
The GGR transform fixes the coefficient of the Wightman function. For
the anti-de Sitter space it is shown that
a reconstruction formula exists for the case of time-like boundary as well.
The restriction on the domain of integration on the boundary is derived.
As a special case, we point out that the formula is valid for 
the BTZ black hole as well.
\end{abstract}
\setcounter{page}{0}
\thispagestyle{empty}
\clearpage
\section{Introduction}
Holography is a duality transformation relating
a pair of field theories, one living in some manifold and the other
on its boundary, suitably defined. An extensively
studied example of holographic duality is the AdS-CFT
correspondence. It relates a theory of closed strings at weak coupling on
the product of a five-dimensional sphere and a five-dimensional anti-de 
Sitter space to
a gauge theory of three-branes on the conformal
boundary of the latter. 
The converse problem of bulk reconstruction, which we deal with here,
attempts to directly obtain a field in the bulk of the manifold from one on
the boundary, usually as an integral over a portion of the boundary through
a kernel. Such relations have been obtained for
manifolds with  constant curvature 
\cite{Balasubramanian:1999ri,Bena:1999jv,Hamilton:2005ju,Hamilton:2006az,
Hamilton:2006fh, Kabat:2012hp,Kabat:2013wga,Kabat:2015swa,Roy:2015pga,
Kabat:2017mun,Sanches:2017xhn,Goto:2017olq,Roy:2017hcp,Sarosi:2017rsq}. 
Determination of functions and distributions on a manifold from the knowledge of distributions on a suitable class of submanifolds is the subject of study in integral geometry. 
This entails specifying appropriate classes of functions on the manifold and on the submanifolds and relating those through integral transforms. 
In the present article, we consider scalar
fields on manifolds of constant curvature,
namely, the de Sitter and the anti-de Sitter spaces. Using integral
geometric techniques of horospherical transform we relate such fields to
the ones on the boundary. In particular, the fields in the bulk of these
spaces are expressed as the inverse of a horospherical transform, called the
Gel'fand-Graev-Radon (GGR) transform
\cite{gel1966generalized,helgason}. 
The present article generalizes similar computations in the Euclidean anti-de
Sitter space \cite{brs}.
Generalization to the two-dimensional hyperbolic manifold over local
fields has been worked out too \cite{BR}.

The relation between the bulk and boundary fields
in the anti-de Sitter
space is given by the HKLL formula {\cite{Hamilton:2005ju}}. 
It expresses the bulk field in the anti-de Sitter space as an integral of the boundary field with a kernel. 
The domain of integration is chosen to be a space-like region of the boundary.
We find that interpreted as the integral transform the formula is valid in
odd dimensions, the kernel being plagued with discontinuity of
coefficients in even dimensions. We also show that in odd dimensions
the inverse GGR transform allows for a similar formula
with the time-like portion of the boundary as the domain of integration. 
The restriction of the domain of integration on the boundary is derived as a
result of consistency of change of variable.
We also establish a 
similar formula for the odd-dimensional de Sitter spaces,
although, as is well-known, the time dependence of the field theories somewhat
obscure the nature of holography on a de Sitter space
\cite{Strominger:2001pn,Hull:1998vg,Hull:1998ym,Balasubramanian:2001rb,
Witten:2001kn,Bousso:2001mw,Balasubramanian:2001nb,Klemm:2001ea,
Balasubramanian:2002zh}. 
As a check on the consistency of the results we evaluate the two-point
correlation function for the scalars in the bulk exactly. The two-point
function is expressed in the terms of a Gauss hypergeometric function,
thereby yielding the Wightman function. The coefficient of the correlation
function is fixed by the structure of the GGR transform. 
Finally, as a special case, we recall 
that the three-dimensional anti-de Sitter space can be identified with the
group manifold of $SL(2,\R)$, a quotient of which is the BTZ black hole 
\cite{keski}. Through an appropriate identification of coordinates we 
demonstrate 
that the bulk reconstruction formula is also valid for the bulk
of the BTZ black hole. 
The strategy to derive the bulk reconstruction formula is the same as the one 
employed earlier \cite{brs, BR}. We restrict our attention to scalar fields.
The $n$-dimensional 
de Sitter and anti-de Sitter spaces, referred to as the bulk, are 
presented as quadrics in a $(n+1)$-dimensional flat space, referred to as
the embedding space, with a
metric of appropriate indefinite signature.
A linear equation
in terms of the coordinates of the embedding space and its light cone defines a horosphere.
The GGR transform of a field in the bulk gives a field on the horosphere. The inverse gives a field in the bulk. By identifying the conformal
boundary within the
horosphere we show that if the field possesses certain scaling properties on the light cone, then the kernel transforming it into the bulk can be defined
through an integral over a portion of the boundary. The kernel comes with a
constant coefficient depending on the dimension of the bulk as well as the
scaling dimension of the scalar field on the light cone. Part of it is fixed
by demanding consistency of the GGR transform and its inverse. The
coefficient of the inverse GGR transform is usually singular for certain
dimensions. This originates in the well-known ill-posedness of the inverse Radon
transform. However, combined with singular terms arising from the scaling behavior of the field,
the coefficient of the Kernel turns out to be non-singular for odd
dimensions, but for a volume factor of hyperbolic spaces, 
which is to be understood in a regularized sense in each case.

In the following two sections we obtain the bulk scalar fields from the boundary using the inverse GGR transform for de Sitter and anti-de Sitter spaces,
respectively. In both cases, the coefficient of the kernel, 
apart from the volume factor,
is continuous and non-singular only in odd dimensions. Furthermore, in the
anti-de Sitter space, two cases arise. The domain of integration, that is, the
domain of influence on the boundary may be either spacelike or time-like. 
The coefficients are different in the two cases.
Evaluation of the inverse GGR transform requires using Dirac
delta distributions in spaces with metrics of non-Euclidean signature 
\cite{gel1966generalized}. We include this computation and some relevant
integrals in two appendices. 
\section{de Sitter space}
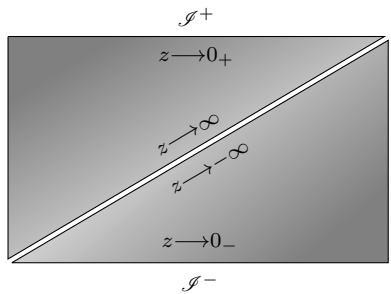
\begin{figure}[h]
  \centering
  \begin{tikzpicture}
  \shadedraw [shading angle=235] (0.05,0) -- (5,0) node[midway,below] {$\scriptstyle\mathscr{I}^-$} node[midway,above] {$\scriptstyle z\rt 0_-$} -- (5,2.95) 
-- node[midway,below, sloped] {$\scriptstyle z\rt-\infty$}  (0.05,0);
  \shadedraw [shading angle=45] (0,3) -- (4.95,3) node[midway,above] {$\scriptstyle\mathscr{I}^+$} 
node[midway,below] {$\scriptstyle z\rt 0_+$} --
(0,.05) node[midway,above, sloped] {$\scriptstyle z\rt\infty$} 
-- (0,3);
\end{tikzpicture}
\caption{Boundaries of the de Sitter space}
\label{fig1}
\end{figure}

The $n$-dimensional de Sitter space, 
to be denoted \ds, is a hyperbolic manifold with a
constant positive curvature. We consider the realization of \ds\
as a quadric in the flat Minkowski space $(\R^{(1,n)},\eta)$ with coordinates
$\{X^a\in\R; a=0,1,\cdots, n\}$ and metric 
$\eta_{ab}=
\left(\begin{smallmatrix}-1&0\\0&{I}_n\end{smallmatrix}\right)$, 
where ${I}_n$ denotes the $n\times n$ identity matrix.
Thus, 
\begin{equation}
\label{eq:ds}
\ds = \{X^a\in\R|\sum_{a,b=0}^n\eta_{ab}X^aX^b=1\}.
\end{equation} 
The light cone $\C_n$ in $(\R^{(1,n)},\eta)$, 
is the set of null vectors $\xi$, 
\begin{equation}
\label{coneds}
\C_n=\{\xi^a\in\R|\sum_{a,b=0}^n\eta_{ab}\xi^a\xi^b=0\}.
\end{equation} 
The region of the light cone with $\xi^0\geqslant 0$ is called 
the positive light cone, denoted $\C_n^+$. 
The metric on $\ds$ is the metric obtained by restriction from $\eta$. 
Let us consider the affine chart $\{(z,x);z\in\R,x\in\R^n\}$ on $\ds$,
such that  
\begin{gather}
\label{Xparads}
X^0 = \frac{z}{2} \left(1-\frac{1+x^2}{z^2}\right), \quad
X^i = \frac{x^i}{z}, \quad
X^n = \frac{z}{2}\left(1+\frac{1-x^2}{z^2}\right), \\
x^2 = \sum_{i=1}^{n-1} (x^i)^2,
\end{gather}
where $x^i$,  denotes a component of $x$.
The metric on \ds\ in this chart is given by
\begin{equation}
\label{ds:metric}
ds^2 = \frac{1}{z^2}\big(-dz^2+\sum_{i=1}^{n-1}(dx^i)^2\big).
\end{equation} 
The coordinates $x$ are spacelike, while $z$ is time-like. 
The volume element of \ds\ is
\begin{equation}
\label{ds:vol}
dV = \frac{1}{z^n}dzd^{n-1}x,
\end{equation} 
where $d^kx$ denotes the volume element of the $k$-dimensional affine
Euclidean space $\R^k$.
The light cone $\C_n$ is a metric cone
$\R_+\times_{\xi^0}\mathbf{S}^{n-1}$ over the
$(n-1)$-dimensional sphere $\mathbf{S}^{n-1}$. The affine coordinates 
on the light cone commensurate with \eq{Xparads} are
\begin{gather}
\label{xiparads}
\xi^i = -\frac{2 \tilde x^i}{1+\tilde{x}^2} \xi^0, 
\quad\xi^n = - \left(\frac{1-\tilde{x}^2}{1+\tilde{x}^2}\right)\xi^0,\quad 
-\infty <\xi^0 < \infty \\
\tilde{x}^2= \sum_{i=1}^{n-1} (\tilde{x}^i)^2,\quad
-\infty < \tilde{x}^i < \infty,\quad i=1,\cdots,n-1.
\end{gather}
In this chart the volume element on the  light cone is 
\begin{equation}
\label{dxids}
\begin{split}
d\xi &= \frac{d\xi^0\cdots d\xi^{n-1}}{\xi^n}\\
&=\frac{2^{n-1} (-\xi^0)^{n-2}}{\left(1+\tilde{x}^2\right)^{n-1}}
d\xi^0 \, d^{n-1} \tilde x
\end{split}
\end{equation}
The inner product of a vector in the de Sitter space and one on the light
cone in this chart is given by 
 \begin{equation}
\label{xidotX:ds}
\xi\cdot X=\sum_{a,b=0}^n\eta_{ab}\xi^a X^b=
\frac{\xi^0\big(-z^2+(x-\tilde x)^2\big)}{z \left(1+\tilde x^2\right)}
\end{equation}
The conformal boundary is at $\xi\cdot X=0$. 
It is situated at $z\rt 0_{\pm}$ and $x\rt\tilde{x}$ in the affine chart. 
The future and past spacelike boundaries are denoted
$\mathscr{I}^{\pm}$, corresponding to
$0 \leqslant z < \infty$ and 
$-\infty < z \leqslant 0$, respectively, as sketched in Figure~\ref{fig1}. 
We present expressions for the former case, the latter being similar.  
\subsection{GGR transform}
\label{GGR:ds}
Let us  consider the horospherical
GGR transform of functions on the de Sitter space. 
The horosphere is given
by the hypersurface
\begin{equation}
\label{Hods}
|\xi\cdot X| -1 =0.
\end{equation}
Let us point out that, 
the modulus, which was not required in the defining equation 
of the horosphere for the Euclidean case \cite{brs}
arises as unlike 
Euclidean anti-de Sitter space, the de Sitter space does not split into two 
disjoint components.
The GGR transform of an integrable function $f(X)$ on \ds\ 
is defined to be \cite{gel1966generalized}
\begin{equation}
 \label{Radonds}
 h(\xi)=\int_{\ds} f(X) \delta\left(|\xi\cdot X|-1\right) dV,
\end{equation}
where the integration is with respect to \eq{ds:vol}.
The inverse of the GGR transform is then given by 
\begin{equation}
 \label{iradds}
f(X) = c_n\int_{\C_n} 
\frac{h(\xi)}{\left(|\xi\cdot X|- 1\right)_+^n} d\xi,
\end{equation}
where we have used the abbreviation
$x_+^a = \theta(x) x^a$, with $\theta$ denoting the Heaviside step function.
Here $c_n$ is a constant which depends on the dimension of the de Sitter space.
To determine the constant we use \eq{Radonds} and \eq{iradds} in conjunction
to obtain 
\begin{equation}
\label{ds-delta}
c_n\mathcal{I} = \delta_{\text{$\ds$}}(X-Y),
\end{equation} 
where we have defined
\begin{equation}
\mathcal{I}=\int_{ \C_n} \frac{\delta\left(|\xi\cdot Y|-1\right)}{%
\left(|\xi\cdot X|- 1\right)_+^n} d\xi,
\end{equation} 
and $\delta_{\text{$\ds$}}(X-Y)$ denotes the Dirac distribution
on $\ds$. Performing the integration and incorporating 
the strength of the Dirac 
distribution \eq{c:ds} fixes $c_n$. Let us describe the computations 
in some detail. 

In order to evaluate the integral $\mathcal{I}$
we choose, without loss of generality, two points $X$ and $Y$ of the de
Sitter space to be
$Y=(1,0)$, $X=(z, 0)$ using the rotational
symmetry of \ds. This is achieved in two steps, fixing $Y$ in the first 
step and then 
using the isotropy subgroup of it to fix  $X$ in the next.
This corresponds to choosing $z=1,x^i=0$ for $Y$ and $x^i=0$
for $X$ in \eq{Xparads}. 
Using \eq{xidotX:ds} the integral $\mathcal{I}$ then simplifies to 
\begin{equation}
\mathcal{I} = \int_{\C_n}
\frac{\delta\left(\left|\tfrac{\xi^0(\tilde{x}^2-1)}{\tilde{x}^2+1}\right|
-1\right)}{
\left(\left|\tfrac{\xi^0(\tilde{x}^2-z^2)}{z(\tilde{x}^2+1)}\right|-1\right)_+^n} d\xi.
\end{equation} 
Inserting \eq{dxids} and defining a new variable
$\rho=\tfrac{\xi^0(\tilde{x}^2-1)}{\tilde{x}^2+1}$,
we express $\mathcal{I}$ as a sum of two integrals, over the domains
$\rho<0$ and $\rho>0$. Integrating over $\rho$ then yields
\begin{equation}
\mathcal{I} = 
(-1)^{n}(2)^{n-1} V_{n-2}\int_0^{\infty}\frac{R^{n-2}dR}{(R^2-1)^{n-1}}
\left(
\frac{1}{\left(\left|\frac{R^2-z^2}{z(R^2-1)}\right|-1\right)_+^n}
+\frac{(-1)^n}{\left(\left|\frac{R^2-z^2}{z(1-R^2)}\right|-1\right)_+^n}
\right),
\end{equation} 
where we have defined the positive number $R$ by $R^2=\tilde{x}^2$ and 
denoted by $V_k$ the volume of the $k$-dimensional unit sphere. 
Changing variable again, to $y=\tfrac{(R^2-z^2)}{z(R^2-1)}$, we 
note that, we have $(|y|-1)_+^n$  and $(|-y|-1)_+^n$ in the two terms of the
integrand. In the domain of $z$ we have chosen, namely $0\leqslant z <
\infty$, we have $y>0$. Thus, $|y|=|-y|=y$. The integral then assumes
the form
\begin{equation}
\mathcal{I} = (-1)^{n}(2)^{n-1}V_{n-2}e^{in\pi/2}\cos\frac{n\pi}{2}
\int_{z}^{\frac{1}{z}}
\frac{z(yz-z^2)^{(n-3)/2} (yz-1)^{(n-3)/2}}{(1-z^2)^{n-2}}
\frac{1}{(y-1)_+^n} dy.
\end{equation} 
The limits of integration vary depending on whether $z$ is greater or less
than unity. To see this we change the variable of integration once again to
$w=yz$. The Integral becomes
\begin{equation}
\mathcal{I}=(-2)^{n-1}V_{n-2}\frac{e^{in\pi/2}\cos\frac{n\pi}{2}}{
(1-z^2)^{n-2}}
\int_{z^2}^1\frac{(w-z^2)^{(n-3)/2}(w-1)^{(n-3)/2}}{(w-z)_+^n}z^ndw.
\end{equation} 
Due to the factor $(w-z)_+^n$ in the denominator of the integrand the integral
is to be interpreted as
\begin{equation}
\int_{z^2}^1 dw = \begin{cases}
\int_z^1 dw,\quad\text{if}\ 0\leqslant z\leqslant 1\\
-\int_z^{z^2} dw,\quad\text{if}\ z\geqslant 1. 
\end{cases}
\end{equation} 
Defining a new variable $t$ as $w=t+(1-t)z$ in the former case and
$w=tz^2+(1-t)z$ for the latter, we arrive at 
\begin{equation}
\label{I:ds}
\mathcal{I}=(-1)^{(n+1)/2}2^{n-1}V_{n-2} e^{in\pi/2}\cos\frac{n\pi}{2}\times
\begin{cases}
\frac{(-1)^{n+1}z^n}{(1+z)^{n-2}(1-z)^n}
&\int\limits_0^1
(1-t)^{(n-3)/2}(t+z)^{(n-3)/2}\frac{dt}{t^n},\\&\qquad\text{if}\ 0\leqslant
z\leqslant 1\\
\frac{z^{(n-1)/2}}{(1+z)^{n-2}(z-1)^n}&\int\limits_0^1
(1-t)^{(n-3)/2}(tz+1)^{(n-3)/2}\frac{dt}{t^n},\\&\qquad\text{if}\ 
z\geqslant 1.
\end{cases}
\end{equation} 
The distance between the points $X$ and $Y$, chosen as above, is   
$(X-Y)^2=-(z-1)^2/z$ with respect to the metric \eq{ds:metric}. 
The constant $c_n$ is given by the inverse of the
coefficient of $|z-1|^n$ in $\mathcal{I}$ evaluated at $z=1$. 
However, the above expression for $\mathcal{I}$ 
shows that the coefficient as $z\rt 1_{\pm}$ match only
when $n$ is odd. In odd dimensions, the constant $c_n$ is given by 
\begin{equation}
\label{cninv}
\begin{split}
c_n&=\frac{1}{c \big(|z-1|^n\mathcal{I})_{z=1}}\\
&=\frac{e^{-i\pi(n+1/2)}\tan\frac{n\pi}{2}}{2^{n-2}\pi V_{n-2}^2}
\frac{\G{n}}{\G{(n-1)/2}^2},
\end{split}
\end{equation} 
where the strength $c$ of $\delta_{\ds}(X-Y)$ is obtained in \eq{c:ds}.
Let us emphasize that the singular $\G{0}$ factors cancelled between 
\eq{c:ds} and \eq{t:int}.
\subsection{Bulk reconstruction}
Assuming that the GGR transform and its inverse are valid for fields we apply 
the considerations of the previous subsection to fields.
We identify the
function $f(X)$ in \ds\ with the bulk field 
and denote it as $\phi(z,x)=f(X)$.
We define the field on the conformal boundary from $h(\xi)$ as 
$\tilde{\phi}(\tilde{x})=h(\xi)$.
We further assume, that on the horosphere \eq{Hods} the boundary 
field scales as 
\begin{equation}
\label{conf:Dds}
\tilde\phi(\lambda\tilde{x}) = \lambda^{-\Delta}\tilde\phi(\tilde{x}),
\end{equation} 
for any function $\lambda=\lambda(\tilde{x})$. 
In particular, this implies
\begin{equation}
\label{cft:assumds}
\begin{split}
h(\xi) &= h(\xi^0,\cdots,\xi^{n-1})\\
&=\tilde{\phi}\left(
\frac{2 \tilde x^1}{1+\tilde{x}^2} \xi^0, 
\frac{2 \tilde x^2}{1+\tilde{x}^2} \xi^0, \cdots,
\frac{2 \tilde x^{n-2}}{1+\tilde{x}^2} \xi^0
\right) \\
&=\left(\frac{2\xi^0}{1+\tilde{x}^2}\right)^{-\Delta} 
\tilde{\phi}(\tilde{x}).
\end{split}
\end{equation} 
Inserting \eq{cft:assumds} and \eq{dxids} in \eq{iradds} yields the bulk
scalar field from $\tilde\phi$ upon integrating over $\xi^0$. In order to
perform the integration over $\xi^0$ we define a new  variable of
integration,
\begin{equation} 
y=\frac{\xi^0}{z(1+\tilde{x}^2)} (-z^2+(x-\tilde{x})^2).
\end{equation} 
From \eq{iradds} we obtain 
\begin{equation}
\label{ds2cone}
\phi(z,x)=
\phi_0(n,\Delta)\int_{\mathscr{I}^\pm} \mathcal{K}(z,x|\tilde x)
\tilde{\phi}(\tilde x) \, d^{n-1}\tilde x \;,
\end{equation}
where the kernel is 
\begin{equation}
\label{smrfuncds}
\mathcal{K}(z,x|\tilde x) = 
{\left(\frac{-z^2+(x-\tilde x)^2}{z}\right)^{\Delta+1-n}}
\end{equation} 
and $\phi_0(n,\Delta)$ is a constant 
\begin{equation}
\phi_0(n,\Delta) = c_n 2^{n-1-\Delta} \int_{-\infty}^{\infty}
\frac{y^{n-2-\Delta}}{\left(|y|-1\right)_+^n} dy.
\end{equation} 
The choice of sign in the kernel guarantees that $y$ does not change sign.
The integral in the expression for $\phi_0(n,\Delta)$ is evaluated as
\begin{equation}
\begin{split}
\int_{-\infty}^{\infty}
\frac{y^{n-2-\Delta}}{\left(|y|-1\right)_+^n} dy &= 
\int_{-\infty}^{-1}
\frac{y^{n-2-\Delta}}{\left(-y-1\right)^n} dy+
\int_{1}^{\infty}
\frac{y^{n-2-\Delta}}{\left(y-1\right)^n} dy \\
&=\big(1+(-1)^{n-\Delta}\big) \int_0^1\frac{y^{\Delta}}{(1-y)^n}dy\\
&=2\pi e^{i\pi(n-\Delta)/2}\frac{\cos\frac{\pi(n-\Delta)}{2}}{\sin n\pi}
\frac{\G{\Delta+1}}{\G{n}\G{\Delta+2-n}}.
\end{split}
\end{equation} 
Using \eq{cninv} this determines the constant $\phi_0$ to be
\begin{equation}
\label{phi0ds}
\phi_0(n,\Delta) = 
\frac{\cos\frac{(n-\Delta)\pi}{2}}{2^{\Delta-1}e^{i\pi(n+\Delta)/2}
V_{n-2}^2\cos^2\frac{n\pi}{2}}
\frac{\G{1+\Delta}}{\G{\Delta+2-n}\G{(n-1)/2}^2}
\end{equation} 
The equation of motion for the bulk field  is obtained from
the action of the Laplacian on $\phi(z,x)$,
\begin{equation}
\label{eomds}
\Box\phi(z,x)=\frac{1}{\sqrt {-g}} \partial_\mu \left( 
\sqrt {-g} g^{\mu\nu} \partial_\nu\phi \right) = m^2\phi(z,x),
\end{equation}
or, expanded using  \eq{ds:metric},
\begin{equation}
\label{eom:full}
z^2\frac{\pa^2\phi(z,x)}{\pa z^2} -(n-2)z\frac{\pa\phi(z,x)}{\pa
z}-z^2\nabla^2\phi(z,x)+m^2\phi(z,x)=0,
\end{equation} 
where $\nabla^2$ denotes the Laplacian with respect to the $x$ coordinates
and the mass of the scalar is given by 
\begin{equation}
\label{massds}
m^2 =  {\Delta(n-1-\Delta)}.
\end{equation}
Thus, there are two modes of the boundary field which correspond to bulk
scalar fields of the same mass. They are related by the exchange 
of $\Delta$ and $n-1-\Delta$ in the above formulas. In the AdS space these
correspond to different asymptotic rates of growth and only one of the modes
is relevant for the boundary scalars. However, in the de Sitter space, if
$m^2 > (\tfrac{n-1}{2})^2$, the two modes are
\begin{equation}
\label{dsDelta}
\Delta_{\pm} =\frac{n-1}{2}\pm i\sqrt{m^2-\left(\frac{n-1}{2}\right)^2},
\end{equation}
with $\Delta_-=n-1-\Delta_+$. Having equal real parts, both modes are to be
included in the boundary theory, as they correspond to equal asymptotic 
growth rates. Reversing the arguments, boundary fields of
scaling dimensions $\Delta$ and $n-1-\Delta$ in \eq{cft:assumds} are to be
used to evaluate bulk quantities.
In particular, both the modes must be included in computing 
correlation functions of scalars of a given  mass $m$
in the bulk.
\subsection{Two-point Correlation function in the bulk}
Let us compute the two-point correlation function of the free scalar field
theory in the bulk similar to the Euclidean anti-de Sitter case treated
earlier \cite{brs} using the expression \eq{ds2cone}. 
Previous estimates \cite{lowe,xiao} 
of this used asymptotic behavior of the kernel
\eq{smrfuncds} and required fixing of coefficients arising from the two modes
by hand. 
Since the coefficient $\phi_0$ is determined by the
inverse Radon transform, the Wightman function is completely determined in
the present approach.

Let us consider the action of a free scalar field of mass $m$ in the bulk,
\begin{equation}
\label{action:bl}
S(\phi) = \int d^{n-1}x dz \sqrt{-g} \left(g^{\mu\nu}\partial_{\mu}\phi(z,x)
\partial_{\nu}\phi(z,x) +m^2 \phi(z,x)^2 \right),
\end{equation} 
where the metric $g$ is given by \eq{ds:metric}.
The generating functional in the presence of a source term
\begin{equation}
S_J(\phi)=\int d^{n-1}x dz \sqrt{-g} J(z,x)\phi(z,x)
\end{equation} 
is given by
\begin{equation}
Z[J]= \int D\phi e^{S(\phi)+S_J(\phi)}.
\end{equation} 
Defining the Radon transform of the source $J$ as 
\begin{equation}
\tilde J(\tilde x) = \phi_0(n,\Delta)
\int  \sqrt{g}\ \mathcal{K}(z,x|\tilde x) J(z,x) dz\ d^{n-1}x
\end{equation} 
and plugging in this along with \eq{ds2cone} in $S+S_J$
we obtain an action for the boundary fields as
\begin{equation} 
\tilde{S}(\tilde{\phi},\tilde{J})=\tilde{S}(\tilde\phi) 
+ \int \tilde J(\tilde x) \tilde \phi(\tilde x) d^{n-1} \tilde x.
\end{equation}
Now that the fields in the bulk and boundary are related by an
\emph{invertible} GGR transform, the actions in the bulk and boundary are
numerically equal,
\begin{equation}
\label{Sbbeq}
S(\phi)+S_J(\phi)=\tilde{S}(\tilde{\phi},\tilde{J}).
\end{equation} 
This relates the generating functional on the boundary
\begin{equation}
\tilde{Z}[\tilde{J}]=\int D\tilde{\phi}
e^{\tilde{S}(\tilde{\phi},\tilde{J})}
\end{equation} 
to that in the bulk, $Z[J]$.
Defining derivatives with respect to the source as
\begin{equation}
\label{J:der}
\frac{\delta}{\delta J'(z,x)} \stackrel{def}{=}
\frac{1}{\sqrt{g}}
\frac{\delta}{\delta J(z,x)} = \phi_0(n,\Delta)
\int d^{n-1}\tilde{x}\  
\mathcal{K}(z,x|\tilde x) \left(\frac{\delta}{\delta \tilde J(\tilde x)}
\right).
\end{equation} 
we can now relate the two-point correlation functions evaluated from $Z[J]$ and
$\tilde{Z}[\tilde{J}]$ as
\begin{equation} 
\begin{split}
\label{BBCor}
\langle \phi(z_1,x_1) \phi(z_2,x_2)\rangle = 
\phi_0(n,\Delta)^2 
\int d^{n-1}\tilde{x}_1 d^{n-1}\tilde{x}_2
\K(z_1,x_1|\tilde{x}_1)\K(z_2,x_2|\tilde{x}_2)
\langle
\tilde{\phi}(\tilde{x}_1)\tilde{\phi}(\tilde{x}_2) 
\rangle
\end{split}
\end{equation} 
All other moments in the presence of polynomial interaction can be similarly
related \cite{brs}.
Let us reiterate that the ensemble average in \eq{BBCor} is justified by the
equality of actions \eq{Sbbeq}, which follows from the invertibility of the
GGR transform. 

In view of the scaling \eq{cft:assumds} of the boundary field and since only
fields of equal scaling dimensions possess non-zero two-point function in a
conformal field theory, we take the two-point correlation function
of the boundary theory to be
\begin{equation}
\langle
\tilde{\phi}(\tilde{x}_1)\tilde{\phi}(\tilde{x}_2) 
\rangle
=\frac{1}{(\tilde{x}_1-\tilde{x}_2)^{2\Delta}}.
\end{equation} 
Plugging this and \eq{smrfuncds} in \eq{BBCor} we then have
the two-point correlation function of scalars at two points $X=(z_1,0)$ and
$Y=(z_2,0)$ in the bulk as
\begin{equation}
\label{def:2pt}
\langle \phi(z_1,0) \phi(z_2,0)\rangle = \phi_0(n,\Delta)^2
\iint\frac{d^{n-1}\tilde{x}_1 d^{n-1}\tilde{x}_2\
z_1^{n-1-\Delta}z_2^{n-1-\Delta}}{
(-z_1^2+\tilde{x}_1^2)^{n-1-\Delta}
(-z_2^2+\tilde{x}_2^2)^{n-1-\Delta}
(\tilde{x}_1-\tilde{x}_2)^{2\Delta}
}
\end{equation} 
We have, as in section~\ref{GGR:ds}, used the rotational symmetry of the de
Sitter space to specialize to these two points, without loss of generality.
The correlation function will be expressed as a function of
the invariant distance between these two points,
\begin{equation}
(X-Y)^2 = -\frac{(z_1-z_2)^2}{z_1z_2}.
\end{equation} 
The integration over $\tilde{x}_1$ is performed by expanding the denominator
through Feynman's trick, using 
\footnote{Let us note that the
variable $\xi$ in this subsection is not related to the light-cone
coordinates in the rest of the article. The variable $\eta$ is also not
related to the variable used earlier.}
\begin{equation}
\label{feyn}
\frac{1}{A^aB^b}=\frac{\G{a+b}}{\G{a}\G{b}}\int_0^1\frac{d\xi\
\xi^{a-1}(1-\xi)^{b-1}}{(\xi A+(1-\xi)B)^{a+b}},
\end{equation} 
completing squares and shifting variables. 
This yields
\begin{equation}
\int\frac{d^{n-1}x_1}{
(-z_1^2+\tilde{x}_1^2)^{n-1-\Delta}
(\tilde{x}_1-\tilde{x}_2)^{2\Delta}
} = \frac{\pi^{(n-1)/2}\G{\tfrac{n-1}{2}}}{\G{\Delta}\G{n-1-\Delta}}
\int_0^1\frac{d\xi\xi^{(n-3-2\Delta)/2}(1-\xi)^{(2\Delta-n-1)/2}}{
\left(\tilde{x}_2^2-\tfrac{z^2}{(1-\xi)}\right)^{(n-1)/2}}.
\end{equation} 
Plugging this in \eq{def:2pt} and repeating the same procedure 
for the integration of $\tilde{x}_2$ we obtain
\begin{equation}
\langle \phi(z_1,0) \phi(z_2,0)\rangle = 
\frac{\pi^{n-1}(-1)^{n-1-\Delta}\phi_0(n,\Delta)^2}{\G{\Delta}\G{n-1-\Delta}}
\int_0^1\!\!\! \int_0^1 
\frac{d\xi\xi^{\tfrac{n-3-2\Delta}{2}}(1-\xi)^{
\tfrac{2\Delta-n-1}{2}}\eta^{n-2-\Delta}
(1-\eta)^{\tfrac{n-3}{2}}}{
\left((1-\eta)z_1^2+\eta(1-\xi)z_2^2\right)^{n-1-\Delta}
}.
\end{equation} 
The factor involving $z_1$ and $z_2$ is then written as a Barnes' integral,
using 
\begin{equation}
\label{Barn}
\frac{1}{(1-z)^a} = \frac{1}{2\pi i}\frac{1}{\G{a}}\int_{-i\infty}^{i\infty}
ds \G{a+s}\G{-s}(-z)^s,\quad |\arg(-z)|<\pi.
\end{equation} 
The integration over $\xi$ and $\eta$ are then performed to obtain
\begin{equation}
\label{I0}
\langle \phi(z_1,0) \phi(z_2,0)\rangle = 
\frac{\pi^{n-1}(-1)^{n-1-\Delta}\G{\tfrac{n-1-2\Delta}{2}}\phi_0(n,\Delta)^2}{
\G{\Delta}\G{n-1-\Delta}^2\G{\tfrac{n-1}{2}}}
\left(\frac{z_2}{z_1}\right)^{n-1-\Delta}\mathcal{I}_0,
\end{equation} 
where
\begin{equation} 
\mathcal{I}_0=\frac{1}{2\pi i}\frac{1}{\G{a}}\int_{-i\infty}^{i\infty}
ds \G{\tfrac{n-1}{2}+s}\G{n-1-\Delta+s}
\G{\Delta-\tfrac{n-1}{2}-s}\G{-s}\left(\frac{z_2}{z_1}\right)^{2s},
\end{equation} 
with $|\arg(z_2^2/z_1^2)|<\pi$.
This Barnes'-type 
integral is evaluated by appropriately closing the contour to include
poles from both factors inside. It is expressed in terms of Gauss
hypergeometric functions as
\begin{equation}
\begin{split}
\frac{\sin\pi(\Delta-\tfrac{n-1}{2})}{\pi}\mathcal{I}_0=&
\frac{\G{\Delta}\G{\tfrac{n-1}{2}}}{\G{1-\tfrac{n-1}{2}-\Delta}}
\left(\frac{z_2^2}{z_1^2}\right)^{\Delta-\tfrac{n-1}{2}}
F\left(\Delta,\tfrac{n-1}{2};1-\tfrac{n-1}{2}+\Delta;\tfrac{z_2^2}{z_1^2}
\right)\\
&-
\frac{\G{n-1-\Delta}\G{\tfrac{n-1}{2}}}{\G{1-\tfrac{n-1}{2}-\Delta}}
F\left(n-1-\Delta,\tfrac{n-1}{2};\tfrac{n+1}{2}-\Delta;\tfrac{z_2^2}{z_1^2}\right).
\end{split} 
\end{equation} 
The hypergeometric functions are expressed using quadratic transformation
formulas \cite{NIST,wolfram}
\begin{equation}
F(a,b;a-b+1;z)=(1-\sqrt{z})^{-2a}F\left(a,a-b+1/2;2a-2b+1;
-\tfrac{4\sqrt{z}}{(1-\sqrt{z})^2}\right),
\end{equation} 
with $z=z_2/z_1$ and $|\arg(z_2^2/z_1^2)|<\pi$ to express 
$\mathcal{I}_0$ as a function of the
invariant distance $(X-Y)^2=-(z_1-z_2)^2/z_1z_2$.
Using the property of the Gamma function,
\begin{equation}
\G{z}\G{1-z}=\frac{\pi}{\sin\pi z}
\end{equation} 
and the duplication formula 
\begin{equation}
\G{z}\G{z+\tfrac{1}{2}}=\sqrt{\pi}2^{1-2z}\G{2z}
\end{equation} 
repeatedly to simplify the coefficients we obtain the integral
$\mathcal{I}_0$
as
\begin{equation}
\begin{split}
\left(\frac{z_2}{z_1}\right)^{n-1-\Delta}\mathcal{I}_0=&
2^{2-n}\sqrt{\pi}\G{\tfrac{n-1}{2}}
\left( 
\frac{\G{\Delta}\G{n-1-2\Delta}}{\G{\tfrac{n}{2}-\Delta}}
\left(-\tfrac{4}{(X-Y)^2}\right)^{\Delta}
F(\Delta,\Delta-\tfrac{n}{2}+1;2\Delta-n+2;\tfrac{4}{(X-Y)^2})
\right.\\
&+
\left. 
\frac{\G{n-1-\Delta}\G{2\Delta-n+1}}{\G{\Delta+1-\tfrac{n}{2}}}
\left(-\tfrac{4}{(X-Y)^2}\right)^{n-1-\Delta}
F(n-1-\Delta,\tfrac{n}{2}-\Delta;n-2\Delta;\tfrac{4}{(X-Y)^2})
\right).
\end{split}
\end{equation} 
This is the form valid at large separation of the points $X$ and $Y$. 
Finally, using the analytic continuation formula 
\begin{equation}
\begin{split}
\frac{\G{a}\G{b}}{\G{c}}F(a,b;c;z) &= 
\frac{\G{a}\G{b-a}}{\G{c-a}}(-z)^{-a} F(a,1-c+a;1-b+a;\tfrac{1}{z}) \\
&+
\frac{\G{b}\G{a-b}}{\G{c-b}}(-z)^{-b}F(b,1-c+b;1-a+b;\tfrac{1}{z}) 
\end{split}
\end{equation} 
in the expression for $\mathcal{I}_0$ and plugging in \eq{I0} we obtain the
expression for the two-point correlation function as
\begin{equation}
\label{2pt:inv}
\langle \phi(z_1,x_1) \phi(z_2,x_2)\rangle_{\Delta} = 
\mu(n,\Delta)
F(\Delta,n-1-\Delta;\tfrac{n}{2};\tfrac{1}{4}(X-Y)^2),
\end{equation} 
where
\begin{equation}
\mu(n,\Delta)=
\frac{\pi^{n-1/2}(-1)^{n-1-\Delta}\G{\tfrac{n-1}{2}-\Delta}
\phi_0(n,\Delta)^2}{
2^{n-2}\G{n-1-\Delta}^2\G{\tfrac{n}{2}}}.
\end{equation} 
We have reinstated the $x$ coordinates in the notation, since 
the expression is in
terms of the invariant distance. 

Let us note that the hypergeometric function is invariant under the exchange
between $\Delta$ and $n-1-\Delta$, while the coefficient of the 
two-point function is not.
This expression is derived but for a single mode corresponding to
\eq{cft:assumds} allowed by \eq{massds}, as indicated by the subscript. 
Incorporation of the other mode
simply alters the coefficient. The full two-point function of scalar fields
in the de Sitter bulk is given by
\begin{equation}
\label{Wtmn}
\langle \phi(z_1,x_1) \phi(z_2,x_2)\rangle = 
(\mu(n,\Delta)+\mu(n,n-1-\Delta))
F(\Delta,n-1-\Delta;\tfrac{n}{2};\tfrac{1}{4}(X-Y)^2).
\end{equation} 
This is the Wightman function, with the coefficient fixed by the GGR
transform.
\section{Anti-de Sitter space}
The $n$-dimensional anti-de Sitter space, to be denoted \ads, 
is a hyperbolic manifold with constant negative curvature. 
As for the de Sitter space,
we consider the realization of the anti-de Sitter space
as a quadric in the flat space
$(\R^{(2,n-1)},g)$ with coordinates
$\{X^a\in\R; a=0,1,\cdots, n\}$ and  metric
$g_{ab}=\left(\begin{smallmatrix}-1&0&0\\0&-1&0\\0&0&{I}_{n-1}
\end{smallmatrix}\right)$. Thus
\begin{equation}
\label{eq:ads}
\ds = \{X^a\in\R|\sum_{a,b=0}^n g_{ab}X^aX^b=-1\}.
\end{equation} 
The light cone in $(\R^{(2,n-1)},g)$ 
is the set of null vectors $\xi$
\begin{equation}
\label{cone}
\C_n=\{\xi^a\in\R|\sum_{a,b=0}^n g_{ab}\xi^a\xi^b=0\}
\end{equation} 
defined with respect to the metric $g$. 
The positive light cone $\C^+$ is the set of null vectors with
$\xi^0\geqslant 0$.
We work with the affine chart  $\{(z,x); z\in\R, x\in\R^n\}$ on \ads, such
that 
\begin{gather}
\label{Xpara}
X^0 = \frac{z}{2}\left(1+\frac{1+{x}^2}{z^2}\right),\quad 
X^{i+1} = \frac{{x}^i}{z},\quad
X^n = \frac{z}{2}\left(1-\frac{1-{x}^2}{z^2}\right),
\end{gather}
where $i=0,1,\cdots,n-2$ and 
\begin{equation}
{x}^2 = {\eta}_{ij}{x}^i{x}^j,
\quad -\infty < {x}^i < \infty. 
\end{equation}
The expressions are similar to \eq{Xparads}, with some 
important difference in certain signs and 
the metric. Here, unlike \eq{Xparads}, the vector $x$ can be either 
spacelike or time-like. We consider both cases.
The metric obtained on \ads\  by restricting the $(n+1)$-dimensional 
flat metric $g$ is 
\begin{equation}
\label{ads:ind}
ds^2 = \frac{1}{z^2}\big(-(dx^0)^2+dz^2+(dx^1)^2+\cdots(dx^{n-2})^2\big).
\end{equation} 
The volume element of \ads\ in this metric is
\begin{equation}
\label{ads:vol}
dV=\frac{1}{z^n}dzd^{n-1}x.
\end{equation} 
As before, we choose  commensurate coordinates on the light cone $\C$ as
\begin{equation} 
\label{xipara}
\xi^{i+1}= \frac{2\tilde x^i}{1+\tilde{x}^2} \xi^0,\quad
\xi^n = - \left(\frac{1-\tilde{x}^2}{1+\tilde{x}^2}\right)\xi^0,
\end{equation} 
with $i=0,1,\cdots,n-2$, and 
\begin{equation}
\tilde{x}^2=\sum_{i,j=0}^{n-2}{\eta}_{ij}\tilde{x}^i\tilde{x}^j.
\end{equation}
The volume element on the positive light cone is
\begin{equation}
\label{dxi:para}
\begin{split}
d\xi=
&= \frac{d\xi^0\cdots d\xi^{n-1}}{\xi^n}\\
&=-\frac{2^{n-1}(\xi^0)^{n-2}}{(1+\tilde{x}^2)^{n-1}} 
d\xi^0 d^{n-1} \tilde{x},
\end{split}
\end{equation} 
with $\xi^0\geqslant 0$ and a definite sign of $1+\tilde{x}^2$. 
Using \eq{Xpara} and \eq{xipara} we obtain
\begin{equation}
\label{xiX}
\xi\cdot X = \sum_{a,b=0}^n g_{ab}\xi^aX^b=
-\frac{\xi^0(z^2+(x-\tilde x)^2)}{z(1+\tilde x^2)}, 
\end{equation} 
where $(x-\tilde{x})^2=\sum_{i,j=0}^{n-2}
\eta_{ij}(x^i-\tilde{x}^i)(x^j-\tilde{x}^j)$.
The conformal boundary of \ads\ is  situated at
$z=0$,  $x^i=\tilde x^i$, corresponding to $\xi\cdot X=0$.
\subsection{GGR transform}
The GGR transform of an integrable function in the anti-de Sitter space 
is defined as the integral
\begin{equation}
\label{radon}
h(\xi)=\int_{\ads} f(X) \delta\left(|\xi\cdot X|-1\right) dV
\end{equation}
that restricts an integrable function $f$ in \ads\
to the horosphere
\begin{equation}
\label{sigma}
|\xi\cdot X|-1 =0. 
\end{equation} 
The inverse transform is given by 
\begin{equation}
\label{irad}
f(X) = c_n\int_{\C_{n}^+} 
\frac{{h}(\xi)}{\left(|\xi\cdot X|-1\right)_+^n} d\xi.
\end{equation}
\label{dxi}
As before, $c_n$ is a constant, dependent on the dimension of \ads, determined
through the consistency of \eq{radon} and \eq{irad}. It is determined by the 
consistency of the GGR transform and its inverse as
\begin{equation}
\label{ads-delta}
c_n\mathcal{I} = \delta_{\ads}(X-Y),
\end{equation}
where $\mathcal{I}$ is now defined as the integral
\begin{equation} 
\mathcal{I}=\int_{\C_n^+} \frac{\delta\left(|\xi\cdot Y|-1\right)}{%
\left(|\xi\cdot X|-1\right)_+^n} d\xi 
\end{equation} 
for two points $X$ and $Y$ in \ads.
and $\delta_{\ads}(X-Y)$ denotes the Dirac distribution
on this component.
In order to determine $c_n$ we choose two points 
$X=(z,x)=(z,0)$ and $Y=(z,x)=(1,0)$ as before. 
Using \eq{xiX} and \eq{dxi:para} the integral becomes
\begin{equation}
\mathcal{I}=
-2^{n-1}\int_{\C_n^+}\frac{\delta\left(|-\xi^0|-1\right)}{
\left(\left|-\frac{\xi^0(z^2+\tilde{x}^2))}{z(1+\tilde{x}^2)}\right|
-1\right)_+^n}
\frac{(\xi^0)^{n-2}}{(1+\tilde{x}^2)^{n-1}}d\xi^0d^{n-1}\tilde{x}
\end{equation} 
The boundary with coordinates $\tilde{x}$ may be either spacelike or
time-like. We deal with the two cases separately.
\subsubsection{Case~I: spacelike boundary, $\tilde{x}^2>0$}
Computations in this case are similar to that in the de Sitter case. 
We express the coordinates $\tilde{x}$ in terms of angular and hyperbolic
coordinates, writing $\tilde{x}^2=R^2$, with $R>0$. The integral is
evaluated
exactly as in the case of de Sitter space
with the successive variables of integration
$y=\tfrac{(z^2+R^2)}{z(1+R^2)}$ and $w=yz$ as before. This yields
\begin{equation}
\label{I:ads1}
\mathcal{I}=-2^{n-2}V_{n-2}\times
\begin{cases}
\frac{(-1)^{n-1}z^n (-1)^{\frac{n-3}{2}}}{(1+z)^{n-2}(1-z)^n}
&\int\limits_0^1
(1-t)^{(n-3)/2}(t+z)^{(n-3)/2}\frac{dt}{t^n},\\&\qquad\text{if}\ 0\leqslant
z\leqslant 1\\
\frac{z^{(n-1)/2}(-1)^{\frac{n-3}{2}}}{(1+z)^{n-2}(z-1)^n}&\int\limits_0^1
(1-t)^{(n-3)/2}(tz+1)^{(n-3)/2}\frac{dt}{t^n},\\&\qquad\text{if}\ 
z \geqslant 1,
\end{cases}
\end{equation} 
where $V_{n-2}$ now denotes the volume of the $(n-2)$-dimensional
hyperboloid.
The integral as a function of $z$ is continuous at $z=1$
only when $n$ is odd. 
\subsubsection{Case~II: time-like boundary, $\tilde{x}^2<0$}
Repeating the same steps as in Case-I, with
$y=\tfrac{z^2-R^2}{z(1-R^2)}$ and $w=-yz$ leads to 
\begin{equation}
\label{I:ads2}
\mathcal{I}=(-1)^{(n+1)/2}2^{n-2}V_{n-2}\times
\begin{cases}
\frac{(-1)^{n+1}z^n}{(1+z)^{n-2}(1-z)^n}
&\int\limits_0^1
(1-t)^{(n-3)/2}(t+z)^{(n-3)/2}\frac{dt}{t^n},\\&\qquad\text{if}\ 0\leqslant
z\leqslant 1\\
\frac{z^{(n-1)/2}}{(1+z)^{n-2}(z-1)^n}&\int\limits_0^1
(1-t)^{(n-3)/2}(tz+1)^{(n-3)/2}\frac{dt}{t^n},\\&\qquad\text{if}\ 
z \geqslant 1,
\end{cases}
\end{equation}
where we have now written $\tilde{x}$ in terms of angular and hyperbolic
coordinates with $\tilde{x}^2=-R^2$. 
The continuity of $\mathcal{I}$ as a function of $z$ again
restricts $n$ to odd numbers only. 
We have, for odd $n$,
\begin{equation}
\label{cnI}
c_n^I= \frac{\cos\frac{n\pi}{2}}{2^{n-{3}}\pi e^{i\pi(n-1)/2} V_{n-2}^2}
\frac{\G{n}}{\G{(n-1)/2}^2}
\end{equation} 
\begin{equation}
\label{cnII}
c_n^{II}= 
\frac{\sin n\pi}{2^{n-{2}}\pi {e^{i\pi(n+1)/2}} V_{n-2}^2}
\frac{\G{n}}{\G{(n-1)/2}^2}
\end{equation} 
in the two cases, using \eq{c:ads1}  and \eq{c:ads2} respectively.
\subsection{Bulk reconstruction}
Let us now use the inverse formula \eq{irad} for bulk reconstruction. 
The strategy for bulk reconstruction is the same as before. 
We assume that $\check{f}(\xi)$ is \eq{irad} has a
conformal symmetry on the null cone with conformal dimension $\Delta$, 
\begin{equation}
\label{cft:assum}
\begin{split}
h(\xi)&=h(\xi^0,\xi^1,\cdots,\xi^n) \\
&=h\left(\xi^0,\frac{2\tilde x^0}{1+\tilde{x}^2}\xi^0, 
\frac{2\tilde x^1}{1+\tilde{x}^2}\xi^0, 
\cdots\frac{2\tilde x^{n-2}}{1+\tilde{x}^2}\xi^0, 
- \left(\frac{1-\tilde{x}^2}{1+\tilde{x}^2}\right)\xi^0\right)\\
&=\left(\frac{2\xi^0}{1+\tilde{x}^2}\right)^{-\Delta} 
\tilde{\phi}(\tilde{x}),
\end{split}
\end{equation} 
where we used \eq{xipara} in the second step and $\tilde{\phi}$ 
is a function of
$\tilde{x}^0,\cdots,\tilde{x}^{n-2}$.
Defining 
\begin{equation}
y=\frac{\xi^0(z^2+(x-\tilde{x})^2)}{z(1+\tilde{x}^2)}
\end{equation} 
and inserting \eq{dxi:para} and \eq{xiX} in \eq{irad} we obtain 
\begin{equation}
\label{fxxx}
f(X) = c_n { 2^{n-1-\Delta}} \int_1^{\infty} 
\frac{y^{n-2-\Delta}dy}{(|y|-1)_+^n}
\int\left(\frac{z}{z^2+(x-\tilde{x})^2}\right)_+^{n-1-\Delta}
\tilde{f}(\tilde{x}) d\tilde{x}.
\end{equation} 
The domain of integration of $y$ does not allow $y$ to vanish. Hence, the
expression $z^2+(x-\tilde{x})^2$ 
must be non-vanishing. We have chosen it to be positive here.
Therefore, the domain of integration of $\tilde{x}$ is bounded by 
$z^2+(x-\tilde{x})^2>0$, consistent with the HKLL formula. 
Had we chosen the opposite sign of
$z^2+(x-\tilde{x})^2$, the expression for $f(X)$ would have changed merely by a
sign. Let us also note that this restriction did not arise in the case of de
Sitter space.

Changing variable from $y$ to $t=1/y$ yields
\begin{equation}
\begin{split}
\int_1^{\infty} 
\frac{y^{n-2-\Delta}dy}{(|y|-1)_+^n}
&=\int_0^1\frac{t^{\Delta}}{(1-t)^n}\\
&=\frac{\G{1+\Delta}\G{1-n}}{\G{2+\Delta-n}}.
\end{split}
\end{equation} 
In view of the sign of $1+\tilde{x}^2$ chosen in the two cases above,
this yields
\begin{equation}
\label{ads:res}
\begin{split}
\phi^{I,II}(z,x)=\phi^{I,II}_0(n,\Delta)\int\limits_{z^2+(x-\tilde{x})^2>0}
\mathcal{K}(z,x|\tilde{x})
\tilde{\phi}(\tilde{x}) d^{n-1}\tilde{x},\\
\end{split}
\end{equation} 
where we have defined $\phi^{I,II}(z,x)=f(X)$ in the two cases, along with
\begin{equation}
\label{phi0:ads1}
\begin{split}
\phi_0^I(n,\Delta) =
\frac{1}{2^{\Delta { - 1}}e^{i\pi(n-1)/2}V_{n-2}^2\sin\frac{n\pi}{2}}
\frac{\G{1+\Delta}}{\G{2+\Delta-n}\G{(n-1)/2}^2}
\end{split}
\end{equation} 
and
\begin{equation}
\label{phi0:ads2}
\begin{split}
\phi_0^{II}(n,\Delta) =
\frac{1}{2^{\Delta { - 1}}{ e^{i\pi(n+1)/2}}V_{n-2}^2}
\frac{\G{1+\Delta}}{\G{2+\Delta-n}\G{(n-1)/2}^2}
\end{split}
\end{equation} 
The kernel is the same in both cases, namely, 
\begin{equation} 
\label{smrfunc}
\mathcal{K}(z,x|\tilde x) = 
{\left(\frac{z^2+(x-\tilde x)^2}{z}\right)_+^{\Delta+1-n}}.
\end{equation} 
The bulk field \eq{ads:res} satisfies the Laplace equation
\begin{equation}
\label{eomads}
\Box\phi(z,x)=\frac{1}{\sqrt {-g}} \partial_\mu \left( 
\sqrt {-g} g^{\mu\nu} \partial_\nu\phi \right) 
= m^2\phi(z,x).
\end{equation}
where
\begin{equation}
\label{massads}
m^2 = \Delta(\Delta-n+1).
\end{equation}
In this case the two modes of the bulk field correspond to 
\begin{equation}
\Delta_{\pm}=\frac{n-1}{2}\pm\sqrt{\left(\frac{n-1}{2}\right)^2+m^2}
\end{equation} 
Only the mode corresponding to $\Delta_+$ asymptotically survives.
The two-point correlation function in the bulk of AdS space 
can be obtained similarly as in the previous case. 
\section{BTZ black hole}
Let us now briefly indicate how the present formulation yields a bulk scalar
field for the BTZ black hole. This is not unexpected, but the choice of chart
in \eq{Xpara} helps bringing it out. The BTZ black hole is a quotient of
$SL(2,\R)$, corresponding to the three-dimensional 
anti-de Sitter space \eq{eq:ads}. In terms
of the coordinates of the embedding space $SL(2,\R)$ is parametrized as a
$2\times 2$ real unimodular matrix
\begin{equation}
g = \begin{pmatrix}
X^1+X^2 & X^3+X^0\\X^3-X^0 & X^1-X^2
\end{pmatrix}\in SL(2,\R).
\end{equation}
Writing 
\begin{gather}
X^0 = \sqrt{\frac{r^2-r_+^2}{r_+^2-r_-^2}}\sinh (r_+ t- r_-\phi),\\
X^1 = \sqrt{\frac{r^2-r_-^2}{r_+^2-r_-^2}}\cosh (-r_- t+ r_+\phi),\\
X^2 = \sqrt{\frac{r^2-r_-^2}{r_+^2-r_-^2}}\sinh (-r_- t+ r_+\phi),\\
X^3 = \sqrt{\frac{r^2-r_+^2}{r_+^2-r_-^2}}\cosh (r_+ t- r_-\phi)
\end{gather} 
the BTZ black hole is given by a quotient corresponding to the periodic
identification of $\phi$ as $\phi =\phi+2\pi$. In these coordinates 
the metric takes the form \cite{keski} 
\begin{equation}
 ds^2 = - \frac{(r^2-r_+^2)(r^2-r_-^2)}{r^2} dt^2 +\frac{r^2}{(r^2-r_+^2)(r^2-r_-^2)} dr^2 +r^2(d\phi - \frac{r_+ r_-}{r^2}dt)^2 .
\end{equation}
The coordinates
$(r,t,\phi)$ are related to the coordinates \eq{Xpara} by
\begin{equation} 
\begin{split}
x^0 &= -\sqrt{\frac{r^2-r_-^2}{r^2-r_+^2}} e^{r_+ t-r_- \phi} 
\cosh(r_+\phi-r_-t), \\
x^1 &= -\sqrt{\frac{r^2-r_-^2}{r^2-r_+^2}} e^{r_+ t-r_- \phi} 
\sinh(r_+\phi-r_-t), \\
z &= -\sqrt{\frac{r_+^2-r_-^2}{r^2-r_+^2}} e^{r_+ t-r_- \phi}.
\end{split}
\end{equation} 
Similar coordinates appear in \cite{bss}.
The periodic change $\phi\mapsto\phi+2\pi$ then corresponds to
\begin{equation}
\begin{pmatrix}
x^0\\x^1\\z
\end{pmatrix}
\mapsto
\begin{pmatrix}
x'^0\\x'^1\\z'
\end{pmatrix} = 
e^{-2\pi r_-}\begin{pmatrix}
\cosh 2\pi r_+ &
\sinh 2\pi r_+ & 0\\
\sinh 2\pi r_+ &
\cosh 2\pi r_+ & 0\\
0&0&1
\end{pmatrix}
\begin{pmatrix}
x^0\\x^1\\z
\end{pmatrix}
\end{equation} 
Inserting these in \eq{ads:res} along with the same boost as $(x^0,x^1)$
for the boundary coordinates $(\tilde{x}^0,\tilde{x}^1)$ in \eq{smrfunc} we
obtain $\phi(z',x')=\phi(z,x)$. We conclude that the bulk reconstruction 
formula \eq{ads:res} is valid for the BTZ black hole as well. 
\section{Summary}
To summarize, we have obtained bulk reconstruction formulas for the de Sitter and anti-de Sitter spaces. 
In both the cases, the strategy is the same as the one employed for the
Euclidean version earlier \cite{brs}. We first identify the conformal boundary
within the horosphere defined in the embedding flat spaces.
The field on the boundary is then interpreted as the GGR transform of a bulk
field and assumed to possess a conformal dimension $\Delta$. The bulk field
is written as an integral with a kernel, which is the same as the smearing
function of the HKLL formula with appropriate signatures of the metric. The
form of the kernel is the same in the HKLL formula, as can be guessed through
dimensional considerations. However, the
coefficients are determined using the paraphernalia of GGR transform. 
The coefficients turn out to be well-defined only when the dimension of the
space is odd. This is in contrast with the 
Euclidean case, in which the formula was valid in all dimensions. 
Ill-posedness of the inversion of the integral transform results in singular
factors in the coefficient. 
We show that as a consequence of the assumption
of conformality of the field on the boundary these singularities are
cancelled in the final formula and that too only in odd dimensions. However,
there is an infinite volume factor of a hyperboloid in the case 
of anti-de Sitter space, 
which is to be understood as a regularized number. 

As a test on the consistency of the reconstruction formula, we compute the
two-point correlation function in the bulk de Sitter space. In this case, if
the mass of the scalar satisfies $m^2>\left(\tfrac{n-1}{2}\right)^2$, then
there are two modes of the scalar near the boundary that contribute to build 
the bulk field, which are to be taken into account. Given that the scaling
dimension of the boundary scalar field is fixed by assumption \eq{cft:assumds} 
in deriving the kernel, we know the two-point correlation function of the
boundary fields. Moreover, since the reconstruction is given as a
\emph{invertible} transform, we can relate the correlation functions of the
boundary conformal theory to correlation functions of a scalar field theory
in the bulk even in the presence of certain interactions and at various loops
in a perturbative manner \cite{brs}. The simplest case of the two-point function of a
free theory in the bulk can be evaluated using the reconstruction formula. We
show that for both the modes the bulk two-point function is expressed as a
Gauss hypergeometric function, symmetric under the exchange of the two modes,
$\Delta$ and $n-1-\Delta$, although with different coefficients. Adding these
we obtain the Wightman function \emph{exactly}, without resorting to fixing
coefficients through asymptotic behavior. The fixed coefficient of the
inverse GGR transform thus fixes the coefficient of the Wightman function. 
This shows the usefulness of our approach of 
looking at the bulk reconstruction as an inverse GGR transform. 

On the anti-de Sitter
space, moreover, we obtain two formulas, \eq{ads:res} depending on whether
the domain of integration on the boundary is spacelike or time-like. 
While the causality issue of the latter is not particularly simple, it
generalizes the HKLL formula. 
Moreover, in the case of anti-de Sitter space the restriction on
the domain of integration on
the boundary present in the HKLL formula is derived by demanding consistency
of change of variable \eq{fxxx}. 
Finally, the BTZ black hole can be written as a quotient of the $SL(2,\R)$ group manifold pertaining to the three-dimensional 
anti-de Sitter space. By relating coordinates, we show
that the bulk reconstruction formula obtained here is valid for the BTZ 
black hole too. 

Another form of Radon transform, namely, the geodesic Radon transform, has
been used to study kinematic spaces \cite{geo1,geo2}. 
This formulation is extremely useful
in the context of  Ryu-Takayanagi analysis. Let us briefly mention the
differences between this approach and the one employed here. In the present
approach the coordinates on the boundary is rather explicit, arising through
the embedding. This has been utilized here to deal with the BTZ black hole.
Such an orbifold analysis will be complicated in the geodesic formulation.
While the present formulation is not particularly useful for
Ryu-Takayanagi type  analyses, it helps in dealing with actions directly
as presented earlier \cite{brs}. 
Moreover, the form of the inversion formula, for example, \eq{ds2cone} along
with \eq{phi0ds} allows writing a
source term in the bulk action in terms of the explicit boundary coordinates
\cite{brs}. Hence, the present formulation
facilitates the comparison of correlation functions of the bulk and boundary
theories. 
Moreover, the present formulation, through the computation of
various normalization factors, brings out the dependence of this analysis on
the dimension of the space-time. 
Finally, the essential requirement of a certain conformal behavior of the
field on the projective light-cone,
appears to be a unique feature of the present approach, in line with
\cite{weinberg,deep}.
However, the connection between the two
formulations of Radon transform is well-known. 
We hope that this formulation will be
useful in revealing the structure of the bulk reconstruction problem. 

\appendix
\section{Dirac distribution}
The Dirac delta distribution on an $n$-dimensional 
de Sitter or anti-de Sitter space is 
defined to be proportional to $1/(X-Y)^n_+$, where $(X-Y)$ denotes the
distance between two points $X$ and $Y$ in the space \cite{gelfand1}. 
In order to fix the constant of proportionality let us define
\begin{equation}
\lim_{\mu\rt n/2}\frac{1}{((X-Y)^2)_+^{\mu}} = c\ \delta(X-Y).
\end{equation} 
The constant is then determined by introducing a test function $\phi(X)$ on
the space and integrating over $X$ as
\begin{equation}
\lim_{\mu\rt n/2}\int dV\frac{1}{((X-Y)^2)_+^{\mu}}\phi(X) = c\phi(Y),
\end{equation} 
For simplicity, we take the test function to be unity and choose $Y$ to
be a special point. We consider the cases of de Sitter and anti-de Sitter
spaces separately.
\subsection{de Sitter space}
We choose $Y=(0,\cdots,1)$ and use the affine parametrization \eq{Xparads}
for $X$. Then 
\begin{equation}
(X-Y)^2=\frac{1}{z}\big(x^2-(z-1)^2)\big)
\end{equation} 
Using the volume element \eq{ds:vol} we have 
\begin{equation}
\int dV\frac{1}{((X-Y)^2)_+^{\mu}} = V_{n-2}\int\frac{z^{\mu-n}R^{n-2}dzdR}{%
\big(R^2-(z-1)^2\big)_+^{\mu}},%
\end{equation} 
where we have written the $x$ coordinates in terms of angular and hyperbolic
coordinates such that the norm $x^2=R^2$ with $R>0$. Changing the  variable
of integration from $z$ to $t=(z-1)/R$ then yields
\begin{equation}
\label{c:ds}
\begin{split}
c&=\lim_{\mu\rt n/2}
\int dV\frac{1}{((X-Y)^2)_+^{\mu}} \\
&=\lim_{\mu\rt n/2 }V_{n-2}
\int_0^{1}\frac{dt}{(1-t^2)^{\mu}} 
\int_0^{\infty}\frac{(1+tR)^{\mu-n}dR}{R^{2\mu-n+1}} \\
&=\lim_{\mu\rt n/2 }V_{n-2}
\int_{0}^1\frac{t^{n-2\mu}dt}{(1-t^2)^{\mu}}
\int_0^{\infty} \frac{(1+\rho)^{\mu-n}d\rho}{\rho^{2\mu-n+1}}\\
&=V_{n-2}\frac{\sqrt{\pi}\ \G{0}\G{1-n/2}}{2\ \G{(3-n)/2}}\\
&=V_{n-2}
\frac{\sqrt{\pi}\ \G{0}\G{(n-1)/2}}{2\tan\left(\frac{n\pi}{2}\right)\G{n/2}},
\end{split}
\end{equation} 
where $V_{n-2}$ denotes the volume of the $(n-2)$-dimensional
unit sphere. The second integral in the second line is facilitated by
performing a further change of variable from $R$ to $\rho=tR$.
\subsection{Anti-de Sitter space}
We choose $Y=(1,\cdots,0)$ and use the parametrization \eq{Xpara} for $X$.
Then 
\begin{equation}
(X-Y)^2 = \frac{1}{z}\big(x^2+(z-1)^2\big). 
\end{equation} 
Using the volume element \eq{ads:vol} we then evaluate the integral to
determine $c$. Two cases arise from the indefinite sign of $x^2$. 
\subsubsection{Case-I: $x^2>0$}
If $x^2= R^2$, with $R>0$, then 
\begin{equation}
\int dV\frac{1}{((X-Y)^2)_+^{\mu}} = V_{n-2}\int\frac{z^{\mu-n}R^{n-2}dzdR}{%
\big((z-1)^2+R^2\big)_+^{\mu}},%
\end{equation} 
where $V_{n-2}$ denotes the unbounded volume of the unit hyperboloid.
Changing variables from $z$ to $t=(z-1)/R$ as before we obtain
\begin{equation}
\label{c:ads1}
\begin{split}
c&=\lim_{\mu\rt n/2}
\int dV\frac{1}{((X-Y)^2)_+^{\mu}} \\
&=\lim_{\mu\rt n/2 }V_{n-2}
\int_0^{\infty}\frac{dt}{(1+t^2)^{\mu}} 
\int_0^{\infty}\frac{(1+tR)^{\mu-n}dR}{R^{2\mu-n+1}} \\
&=\lim_{\mu\rt n/2 }V_{n-2}
\int_{0}^{\infty}\frac{t^{n-2\mu}dt}{(1+t^2)^{\mu}}
\int_0^{\infty} \frac{(1+\rho)^{\mu-n}d\rho}{\rho^{2\mu-n+1}}\\
&=V_{n-2}\frac{\sqrt{\pi}\ \G{0}\G{(n-1)/2}}{2\ \G{n/2}},
\end{split}
\end{equation} 
where $\rho=tR$ in the third line is used.
\subsubsection{Case-II: $x^2<0$}
If $x^2= -R^2$, with $R>0$, then
\begin{equation}
\int dV\frac{1}{((X-Y)^2)_+^{\mu}} 
= V_{n-2}\int\frac{z^{\mu-n}R^{n-2}dzdR}{%
\big((z-1)^2-R^2\big)_+^{\mu}},%
\end{equation}
where $V_{n-2}$ denotes the unbounded volume of the unit hyperboloid.
Changing variables again from $R$ to $t=(z-1)/R$ obtain
\begin{equation}
\label{c:ads2}
\begin{split}
c&=\lim_{\mu\rt n/2}
\int dV\frac{1}{((X-Y)^2)_+^{\mu}} \\
&=\lim_{\mu\rt n/2 }V_{n-2}
\int_1^{\infty}\frac{dt}{(t^2-1)^{\mu}} 
\int_0^{\infty}\frac{(1+tR)^{\mu-n}dR}{R^{2\mu-n+1}} \\
&=\lim_{\mu\rt n/2 }V_{n-2}
\int_{1}^{\infty}\frac{t^{n-2\mu}dt}{(t^2-1)^{\mu}}
\int_0^{\infty} \frac{(1+\rho)^{\mu-n}d\rho}{\rho^{2\mu-n+1}}\\
&=V_{n-2}\frac{\G{0}\G{1-n/2}\G{(n-1)/2}}{2\sqrt{\pi}}\\
&=V_{n-2}
\frac{\sqrt{\pi}\G{0}\G{(n-1)/2}}{2\sin\left(\frac{n\pi}{2}\right)\G{n/2}},
\end{split}
\end{equation}
with $\rho=tR$ in the third line.
\section{Two more integrals}
First, the integral in $\rho$ appearing in \eq{c:ads1}
and \eq{c:ads2} is singular. We evaluate it as
follows. First we substitute $\rho'=1+\rho$ followed by $\tau=1/\rho'$. This
yields
\begin{equation}
\begin{split} 
\lim_{\mu\rt n/2 }
\int_0^{\infty} \frac{(1+\rho)^{\mu-n}d\rho}{\rho^{2\mu-n+1}}
&= 
\lim_{\mu\rt n/2 }
\int_0^1 \tau^{\mu-1} (1-\tau)^{2-2\mu-1}d\tau\\
&=
\lim_{\mu\rt n/2 }
\frac{\G{\mu}\G{n-2\mu}}{\G{n-2\mu+\mu}}\\
&=\G{0}.
\end{split}
\end{equation} 

Next, the integrals  appearing  in \eq{I:ds}, \eq{I:ads1} and \eq{I:ads2}
when evaluated at $z=1$ are singular too.
With $t^2$ substituted with $\tau$, the integrals become
\begin{equation}
\label{t:int}
\begin{split}
\int_0^1 (1-t^2)^{(n-3)/2}\frac{dt}{t^n} 
&= \frac{1}{2} \int_0^1 \tau^{-(n+1)/2}(1-\tau)^{(n-3)/2}d\tau\\
&=\frac{\G{(1-n)/2}\G{(n-1)/2}}{\G{0}}\\
&= \frac{\pi}{2\cos\frac{n\pi}{2}}\frac{1}{\G{0}}.
\end{split}
\end{equation} 
In the formulas above we have written the singular factor $\G{0}$ without
regularization to make the cancellation of singular
factors in the expressions for $c_n$ conspicuous.

\end{document}